\documentclass[final,1p,times]{elsarticle} 
\usepackage{graphicx}
\usepackage{amssymb} 
\usepackage{amsthm} 
\usepackage{lineno}

\journal{Nuclear Physics A} 

\begin{document}

\begin{frontmatter} 

% Your Title - please insert
\title{Charge balancing and the fall off of the ridge}

%% Single author (and collaboration) - please insert
\author[auth1,auth2]{Piotr Bo\.zek}
\author[auth2,auth3]{Wojciech Broniowski}
\address[auth1]{AGH University of Science and Technology, Faculty of Physics and Applied Computer Science, al. Mickiewicza 30, 30-059 Krakow, Poland}
\address[auth2]{Institute of Nuclear Physics PAN, ul. Radzikowskiego 152, 31-342 Krakow, Poland}
\address[auth3]{Institute of Physics,  Jan Kochanowski University, 25-406~Kielce, Poland}

\begin{abstract} 
Two-dimensional  correlation functions in $\Delta\eta-\Delta \phi$  for charged hadrons
emitted in heavy-ion collisions are calculated in event-by-event hydrodynamics. With the Glauber 
model for the initial
density distributions in the transverse plane and elongated density profiles in the 
longitudinal direction, the flow patterns in the azimuthal angle of the two-dimensional correlation function 
are properly reproduced. We show that the additional fall-off of 
the same-side ridge in the longitudinal direction can be explained as an effect of local charge conservation 
at a late stage of the evolution. This additional non-flow effect increases the harmonic flow coefficients for 
the unlike-sign particle pairs. 
\end{abstract} 

\end{frontmatter} % do not changedge%% linenr reviewing process
%\linenumbers

\bibliographystyle{model1a-num-names}

\biboptions{sort&compress}

\section{Introduction}

The collective expansion of dense matter in heavy-ion collisions is determined by the geometry of the fireball.
In the longitudinal direction the density profile is assumed to be approximately boost-invariant for
 central rapidities, while in the transverse plane it is asymmetric. The shape and the size of the fireball fluctuates
event-by-event
\cite{Takahashi:2009na,Alver:2010gr,Broniowski:2009fm}. The dependence of the correlation function on the azimuthal
 angle is dominated by the even and odd harmonic flow components 
\cite{Takahashi:2009na,Luzum:2010sp,Werner:2012xh} and can be reproduced in event-by-event viscous hydrodynamics calculations
\cite{Schenke:2011zz,Qiu:2011hf,Bozek:2012fw}. The shape of the same-side ridge in the longitudinal 
(pseudorapidity $\eta$) 
direction can yield important information on the mechanism of the energy deposition in the early stage of the collision
\cite{Gavin:2008ev,Gelis:2009tg,Dusling:2012ig,Gavin:2011gr,Kapusta:2011gt}. The dominant features observed in the data are
the same-side ($\Delta\phi \simeq 0$) and away-side ($\Delta \phi \simeq \pi$)
 ridge, with the latter one exhibiting 
no dependence on $\Delta \eta$, and the former one showing an increase of the correlation function in 
the form of a same-side peak. This structure is much more pronounced for correlations of the unlike-sign hadrons. 
\cite{Agakishiev:2011pe,Abelev:2008un}.

The formation of charges  at a late stage of the collective  evolution induces strong 
correlations between the unlike-sign hadrons in the pseudorapidity and azimuthal angle
 \cite{Bass:2000az,Jeon:2001ue,Bozek:2004dt,Aggarwal:2010ya}. We argue that the same mechanism explains the observed shape of the 
same-side ridge in the (unbiased) two-dimensional correlation function \cite{Bozek:2012en}. These non-flow correlations 
from the local charge conservation also yield a small contribution to the flow coefficients.

\section{Hydrodynamic model with local charge conservation}

We use a 3+1-dimensional viscous hydrodynamic model \cite{Bozek:2011ua} with bulk and shear viscosities 
\cite{Bozek:2009dw}. The calculations are run event-by-event with the initial entropy density  generated 
as coarse-grained distributions from GLISSANDO \cite{Broniowski:2007nz}. 
Particle emission at freeze-out is performed using the THERMINATOR code \cite{Chojnacki:2011hb}.

\begin{figure}[htbp]
\begin{center}
 \includegraphics[width=0.4\textwidth]{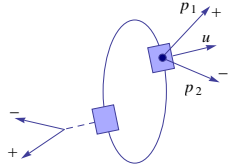}
 
\end{center}
\caption{Mechanism of the generation of charge conservation correlations from resonance decays and local pair creation.}
\label{fig:edu}
\end{figure}

%Local charge conservation forces strong correlation between the charges of emitted particles. 
We implement the local charge conservation in 
 the statistical emission code. Opposite-charge particles are emitted in pairs from the same fluid element. Thus, they feel the same collective flow velocity
which collimates their motion. The spread in their relative momenta comes the thermal 
motion 
(Fig. \ref{fig:edu}). The procedure used to generate  particle distributions presented here
 yields slightly stronger correlations compared to \cite{Bozek:2012en}, including pairs from 
resonance decay cascades as well.
Our hydrodynamic model  reproduces well the particle spectra and flow coefficients.

\section{Two-dimensional correlation functions}

\begin{figure}[htbp]
\begin{center}
 \includegraphics[width=0.49\textwidth]{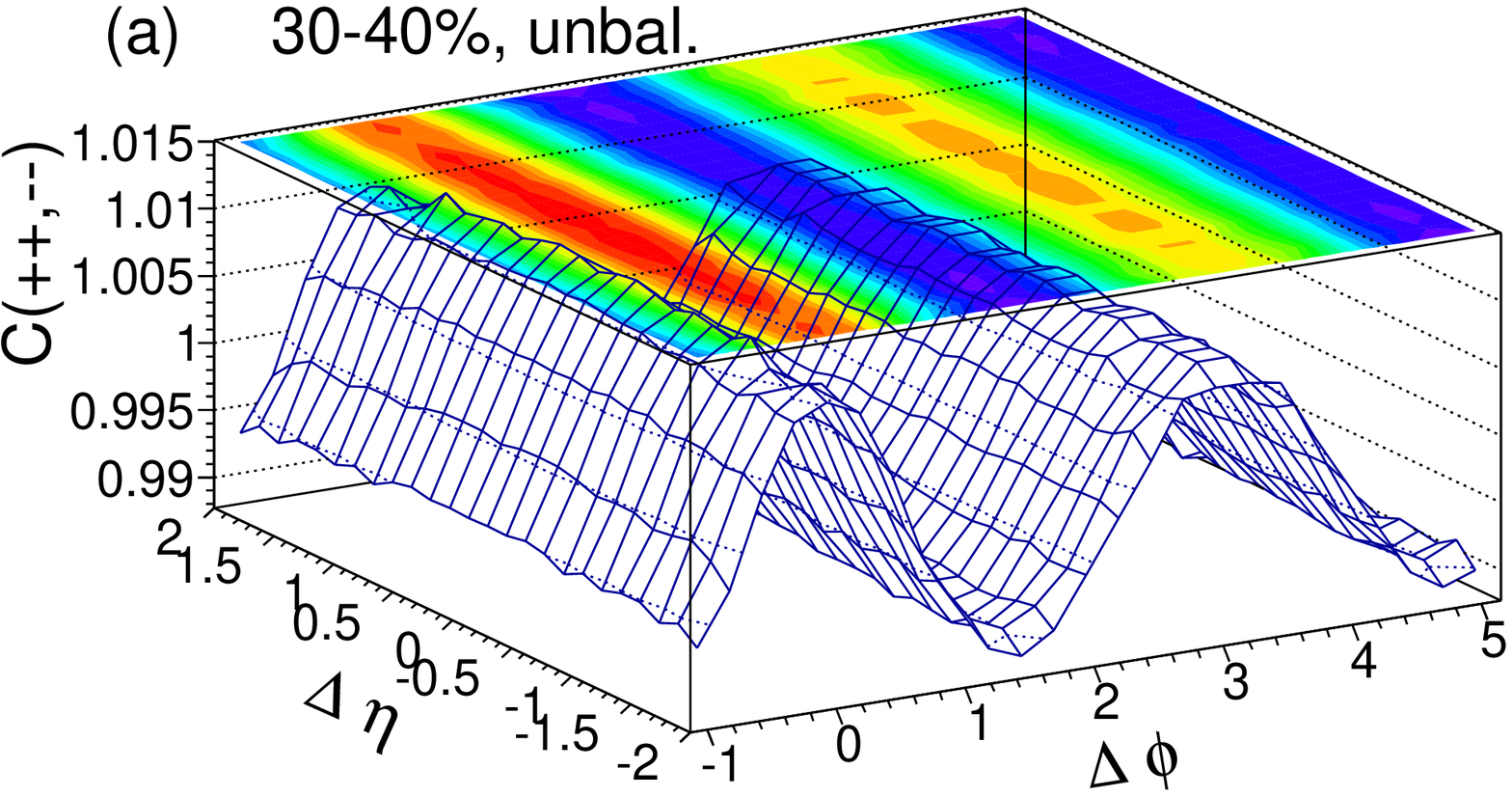}
 \includegraphics[width=0.49\textwidth]{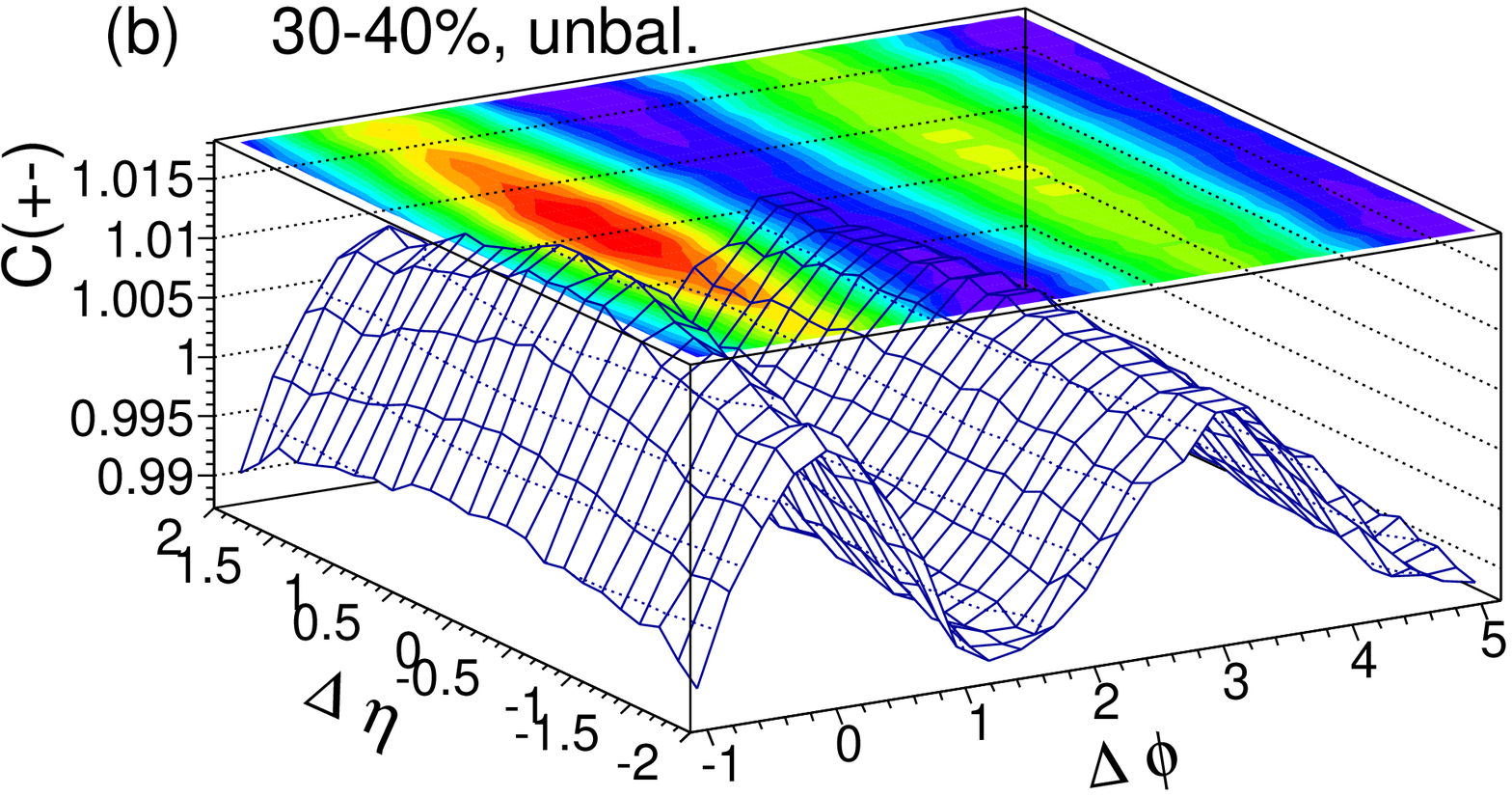}
 \includegraphics[width=0.49\textwidth]{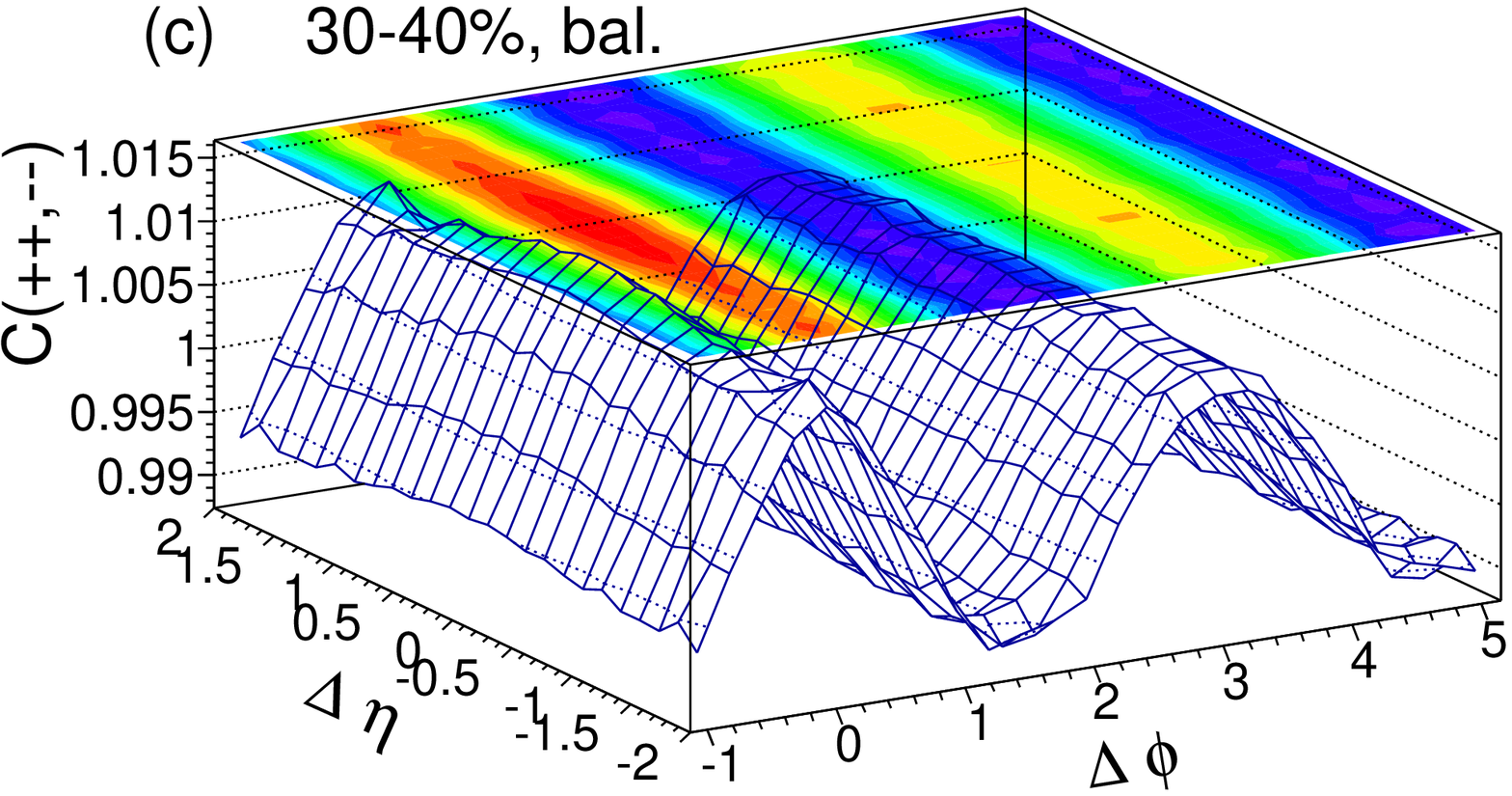}
 \includegraphics[width=0.49\textwidth]{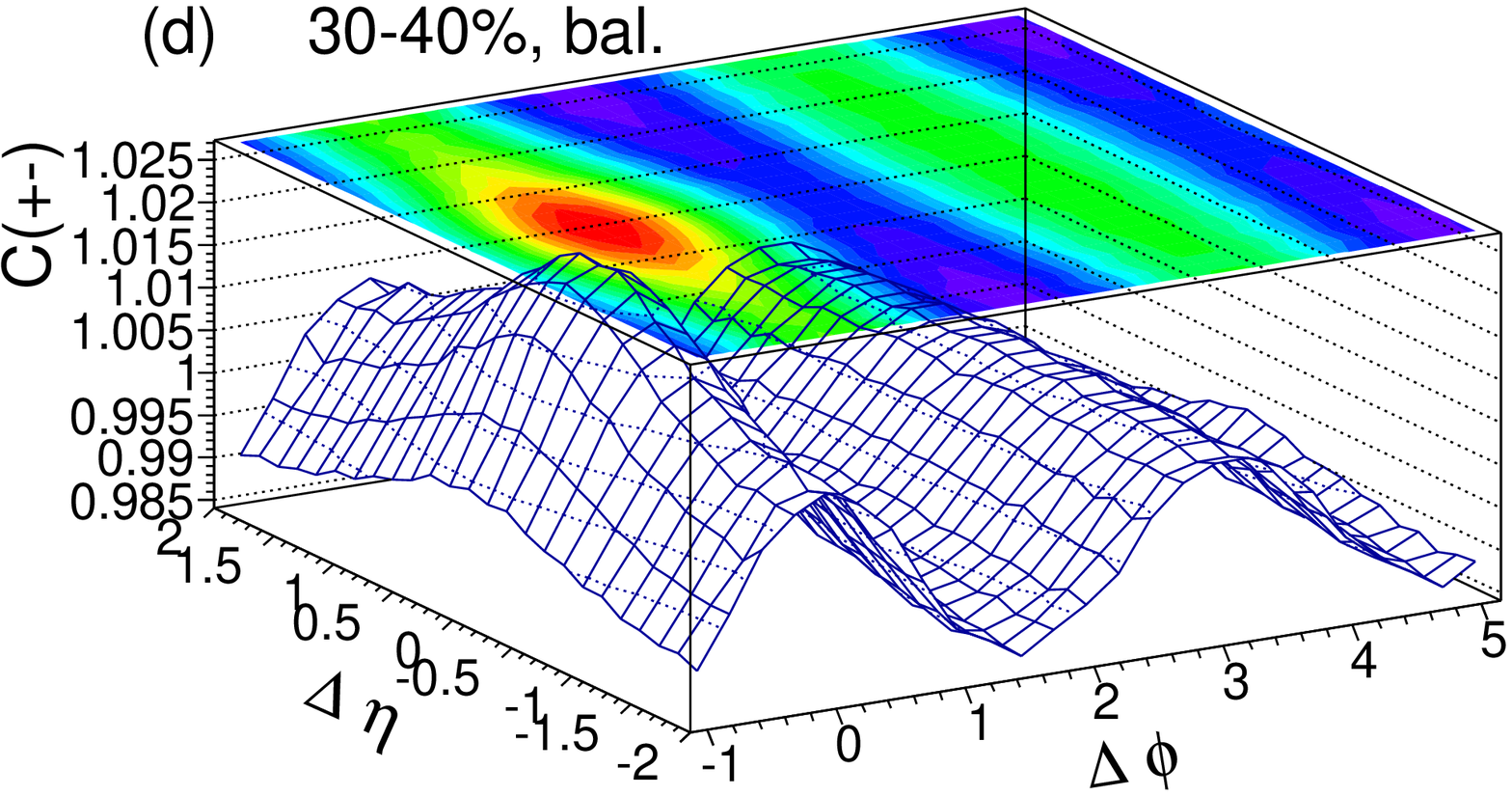}
\end{center}
\vspace{-5mm}
\caption{Two-dimensional correlation functions for particle emission without local charge conservation
 (panels (a) and (b)) and with local charge conservation (panels (c) and (d)), Au-Au collisions at $\sqrt{s_{NN}}=200$~GeV in
 30-40\% centrality class  ($T_f=150$~MeV, $0.2 < p_T < 2$~GeV).}
\label{fig:td}
\end{figure}

Two-dimensional correlation functions
\begin{eqnarray}
C(\Delta \eta, \Delta \phi) = {N^{\rm pair}_{\rm real}(\Delta \eta, \Delta \phi)}/{N^{\rm pair}_{\rm mixed}(\Delta \eta, \Delta \phi)}, \label{eq:def}
\end{eqnarray}
are calculated for the like- and unlike-sign  hadron pairs. In the case of the uncorrelated statistical emission 
of particles at freeze-out there is only a small difference in the results for different charge combinations 
(panels (a) and (b) in Fig.~\ref{fig:td}). 
Some weak short range correlations between unlike charged particles come 
from resonance decays. The same-side peak in the unlike-sign correlation function (panel (b)) is much smaller 
than observed experimentally.

The local charge conservation mechanism generates noticeable correlations
 in the directions of the emitted hadron pairs.
 The unlike-sign particle pairs generated  at freeze-out are collimated by the common collective flow. The effect is clearly seen 
as a strong same-side peak in the unlike-sign correlation function (panel d)) that is not observed for like-sign pairs (panel c)). These features of the same-side ridge are compatible with the experiment \cite{Agakishiev:2011pe}.

\section{Flow coefficients}

\begin{figure}[htbp]
\begin{center}
 \includegraphics[width=0.35\textwidth]{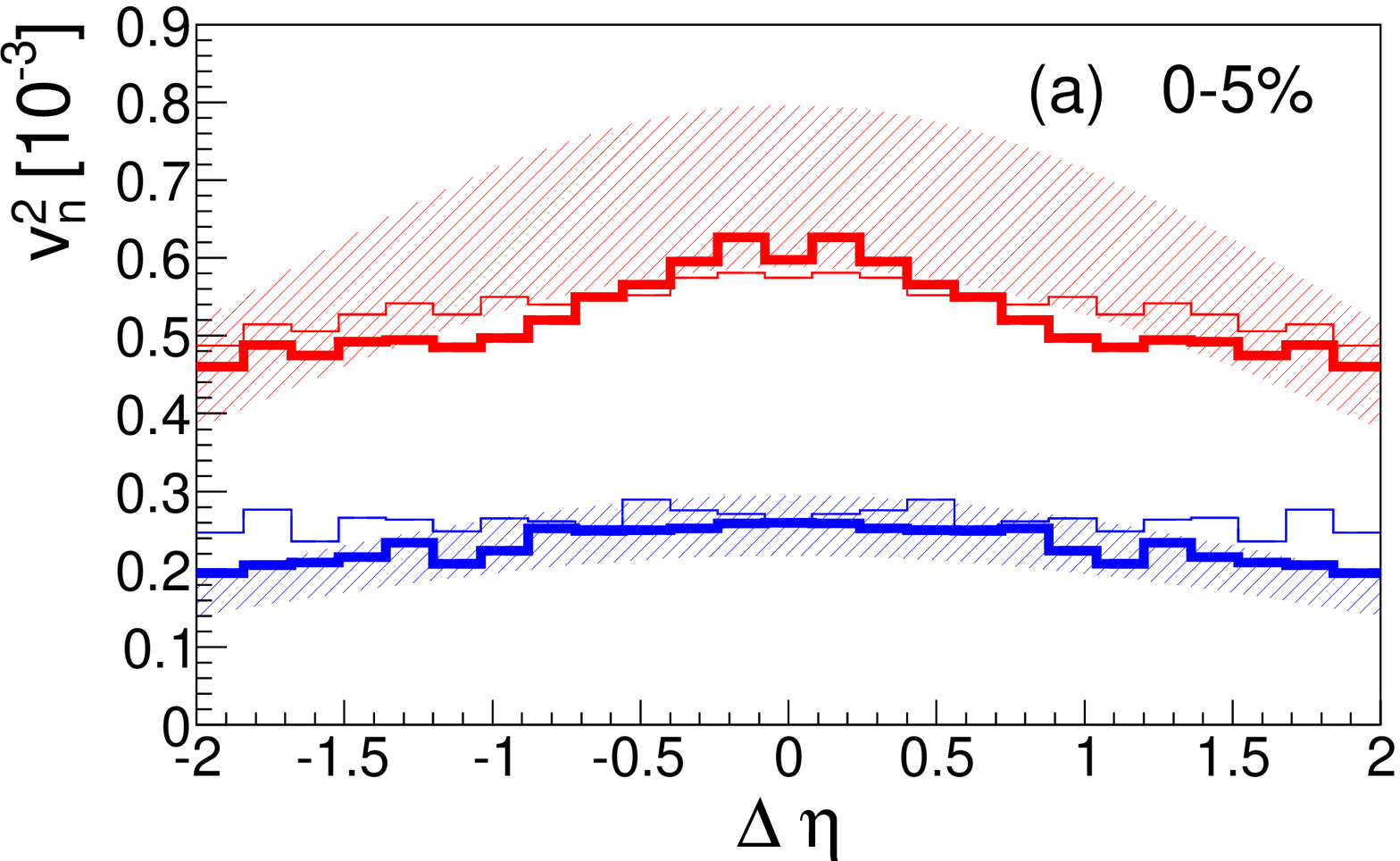}
 \includegraphics[width=0.35\textwidth]{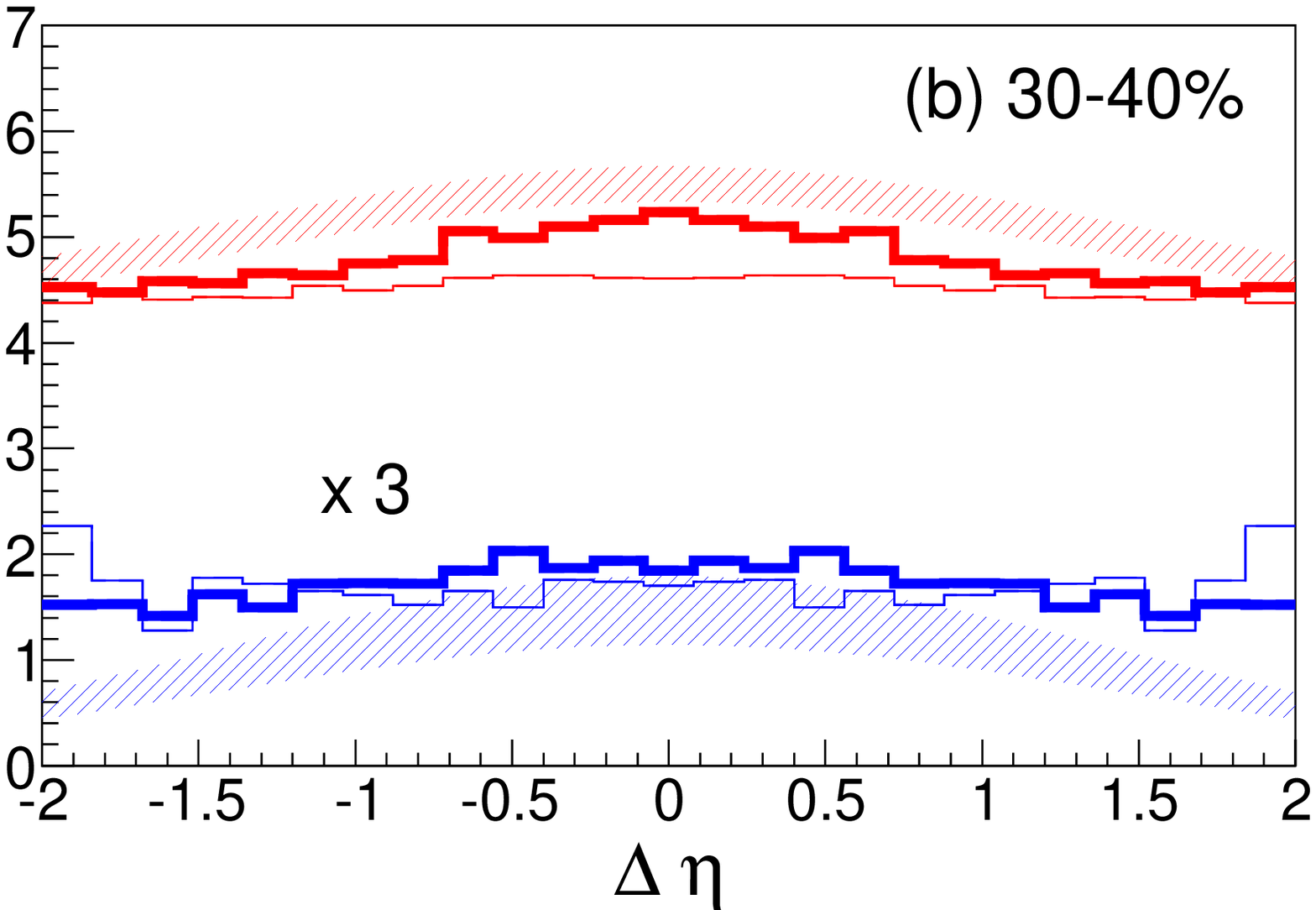}
\end{center}
\vspace{-5mm}
\caption{Elliptic (upper lines)  and triangular (lower lines) flow coefficients plotted as function of the relative 
pseudorapidity of the pair. The thick and thin lines represent the results of the simulations with and without the local charge conservation, respectively 
($T_f=150$~MeV, $0.15 < p_T < 4$~GeV). The shaded bands are extracted from  the measured two-dimensional correlation functions  \cite{Agakishiev:2011pe}. }
\label{fig:vneta}
\end{figure}

The two-dimensional correlation function shown in Fig. \ref{fig:td} contains the information on 
all the harmonic flow 
coefficients. The local charge conservation effect gives a non-flow contribution to the 
observed $v_n^2$ coefficients. The 
charge conservation correlations decrease with the multiplicity and with the 
 pseudorapidity separation of the particle pair
(Fig. \ref{fig:vneta}). 
The non-flow correlation increase $v_n^2$ at small pseudorapidity separations, in a similar 
way as the resonance decays. The mechanism of local charge conservation explains the magnitude and the 
range in $\Delta \eta$ of the non-flow correlations in $v_n$ and is consistent with 
the measured difference of the flow coefficients for unlike and like-sign pairs.

\begin{figure}[htbp]
\begin{center}
 \includegraphics[width=0.4\textwidth]{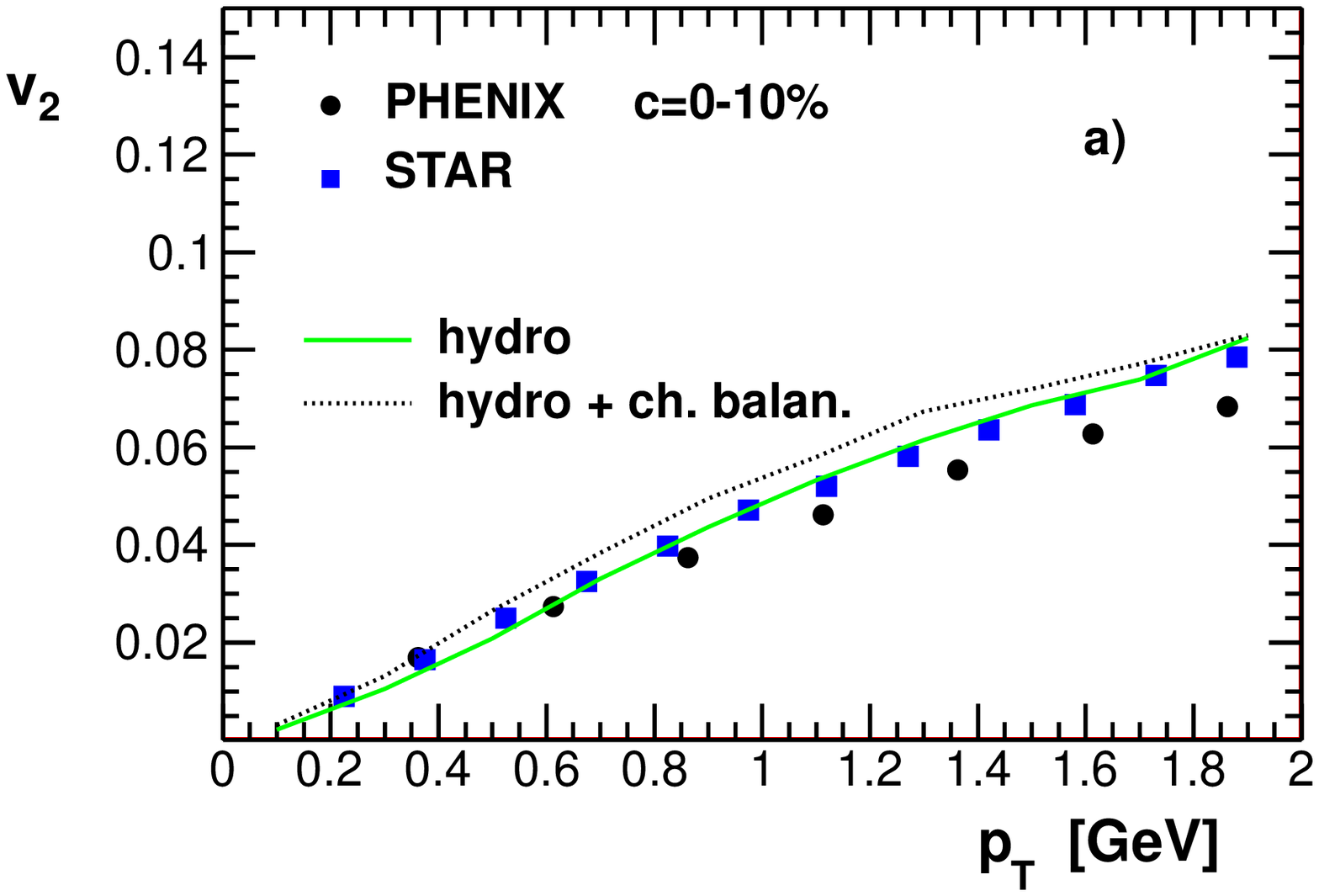}
\includegraphics[width=0.4\textwidth]{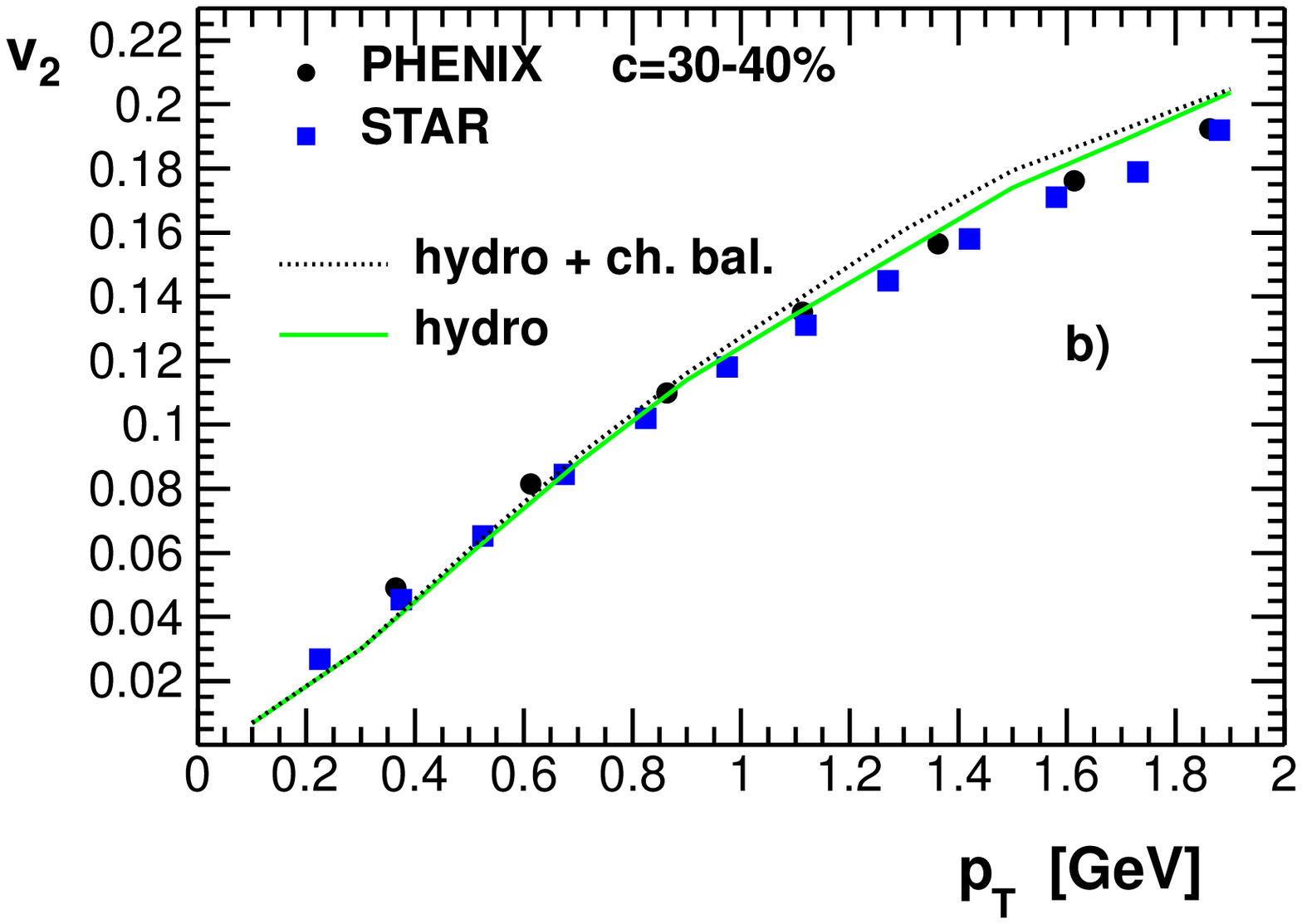}
\end{center}
\vspace{-5mm}
\caption{The elliptic flow coefficient of charged particles  with (dashed lines) or without (solid lines) local charge conservation mechanism, 
centrality 0-10\% (panel a) and 30-40\% (panel b). Data from \cite{Adams:2004bi,Adare:2011tg}.}
\label{fig:vnpt}
\end{figure}

Figure \ref{fig:vnpt} shows the $p_\perp$ dependence of $v_2$. The local charge conservation effects 
give a small contribution to the $v_2$ measured with the second cumulant method. These effects may be reduced
when using a pseudorapidity gap for the pair or a $Q$-vector defined at forward rapidities. 
The mechanism of local charge conservation presents a dominant source of non-flow correlation 
in heavy-ion collisions. It gives a noticeable contribution to $v_1$ and $v_2$. The charge splitting 
induced by the local charge conservation effects in $v_1$ gives a large contribution to charge parity 
violation signals \cite{Schlichting:2010qia,Hori:2012kp}. However, to reproduce the magnitude 
of the charge-independent correlations \cite{Abelev:2009ad}, the total transverse momentum conservation
 has to be imposed  \cite{Bzdak:2012ia}.
Within the hydrodynamic model  with flow and local charge conservation  the 
charge balance functions in relative pseudorapidity can be extracted as well. 
The results are in satisfactory agreement with the data \cite{Aggarwal:2010ya}.

Supported by Polish Ministry of Science and Higher Education, grant N~N202~263438, and National Science 
Centre, grant DEC-2011/01/D/ST2/00772. 

%\section{Summary}

\section*{References}

\bibliography{../../../hydr}

\end{document}